\documentclass[%
reprint,
 amsmath,amssymb,
 aps,
 prl,
]{revtex4-1}

\usepackage{natbib}
\usepackage{graphicx}
\usepackage{dcolumn}
\usepackage{bm}

\begin{document}

\title{A framework for solvation in quantum Monte Carlo}
\author{Kathleen A. Schwarz}
\email{ kas382@cornell.edu}
\affiliation{Department of Chemistry and Chemical Biology, Cornell University}
\author{Ravishankar Sundararaman}
\author{Kendra Letchworth-Weaver}
\author{Tom\'as A. Arias}
\affiliation{Department of Physics, Cornell University}
\author{Richard G. Hennig}
\affiliation{Cornell University Department of Materials Science}

\date{\today}

\begin{abstract}
%
%
Employing a classical density-functional description of liquid
environments, we introduce a rigorous method for the diffusion quantum
Monte Carlo calculation of free energies and thermodynamic averages of
solvated systems that requires {\em neither} thermodynamic sampling
{\em nor} explicit solvent electrons.  We find that this method yields promising results and
small convergence errors for a set of test molecules.  It
is implemented readily and is applicable to a range of challenges
in condensed matter, including the study of transition states of
molecular and surface reactions in liquid environments.
\end{abstract}

\maketitle

The physics of solvation, though poorly understood, plays a critical
role in a wide range of systems from the biological to the
technological.  For example, the pathways in protein folding
\cite{Lazaridis:1997p502, Rhee:2004p503, FernandezEscamilla:2004p501}
and transition states of ionic reactions
\cite{Kim:2009p515,Regan:2002p516} are known to be highly solvent
dependent.  Development of a fundamental understanding of the kinetics
which underlie such processes requires an accurate description of the
quantum mechanical processes involved in bond breaking and formation
in solution.

Unlike less rigorous electronic structure methods, quantum Monte Carlo
methods \cite{Foulkes:2001p332} do provide the required accuracy for
transition states, reactants, and products needed to give reliable
information on reaction pathways. However, a full quantum Monte Carlo
treatment, in principle, requires solving Schr\"odinger's equation for
all of the involved solvent molecules, a difficulty radically
compounded by the need to sample the phase-space of all
thermodynamically relevant configurations of the solvent.  This
{\em Communication} introduces a framework for the treatment of solvation in
diffusion Monte Carlo which completely eliminates the need for
explicit solvent electrons and such phase-space sampling, while remaining
completely \textit{ab initio} and exact in principle --- in the same
sense in which density-functional theory meets these criteria.

Previous attempts to model the effects of solvation in quantum Monte
Carlo fall into two categories: (i) simulation of the environment through
molecular dynamics \cite{Cho:2010p492, Grossman:2005p498}, and
(ii) introduction of a polarizable continuum \cite{Amovilli:2006p491,
  Amovilli:2008p495}.  The former approach, in principle, captures
molecular-scale effects such as solvation shells and non-local
dielectric response.  However, molecular dynamics with empirical
potentials \cite{Cho:2010p492}, though benefiting from a simplified
description of the solvent, depends on a highly parameterized
potential to describe the interactions between molecules.
Additionally, such calculations require phase-space sampling, which
makes the calculations costly --- especially if individual electronic structure calculations for the solute are needed for each fluid configuration.  Approaches have been proposed to mitigate this expense within \textit{ab initio} molecular dynamics calculations \cite{Grossman:2005p498}, but this approach also is extremely expensive due to the need to include of all of the solvent electrons and nuclei.  

In contrast, the polarizable continuum model (PCM) is a
computationally inexpensive tool for approximating solvation energies
without phase-space sampling.  Unfortunately, this model
lacks theoretical justification for the treatment of water as a
continuum on molecular length scales, and so does not constitute a
truly {\em ab initio} method.  Indeed, the pioneering work on
solvation studies within diffusion Monte Carlo electronic structure
methods \cite{Amovilli:2008p495} rested on the \emph{ad hoc}
introduction of a polarizable medium and required a set of spheres
with radii determined empirically.  A further, practical disadvantage
of the aforementioned PCM approach is that it requires potentially costly
statistical evaluation of solvent potentials, and so it has yet to yield
meaningful comparisons between predicted solvation energies and
experiment.

Joint density-functional theory \cite{Petrosyan:2005p493} circumvents both the inaccuracies inherent in the polarizable continuum model and empirical molecular dynamics and the expense associated with \textit{ab initio} molecular dynamics.  This work presents an integrated approach to quantum Monte Carlo calculations within this new, rigorous statistical treatment of the solvent, and introduces a variational theorem which abrogates much of the computational complexity associated with achieving full self-consistency between the solute and the solvent.  While our approach in theory requires no adjustable parameters, we employ a simplified, approximate version of the theory in this first demonstration that does require a {\em single} empirically adjusted parameter. 

This {\em Communication} begins with a brief review of joint
density-functional theory and then describes a
quantum Monte Carlo formulation of that theory, compares predictions to
solvation data, and shows that self-consistency to
within chemical accuracy is achievable with the computational cost of a single quantum Monte Carlo
calculation.

\textit{Joint density-functional theory and quantum Monte Carlo.}
Joint density-functional theory~\cite{Petrosyan:2005p493,Petrosyan:2007} states that, in principle, the exact quantum and thermodynamically averaged electronic and nuclear densities and the exact free energy of a solvated system can be obtained by minimization of a universal free-energy functional, without the need to sample explicitly the phase space of all possible configurations of solvent molecules.  This minimization is carried out over all realizable average solute electron densities $n(r)$ and average solvent site densities $N_\alpha(r)$.  (In the present work, $N_\alpha(r)$ refers to the single-particle densities of the oxygen nuclei and protons comprising the solvent, water.)
Note that $n(r)$ here refers only to the electron density of the
solute because the electron density of the solvent need not be
considered explicitly.  Naturally, the underlying indistinguishability
of electrons implies that there is no unique decomposition of the
total electron density into solute and solvent contributions.
Consequently, the solution to the minimization problem is highly
degenerate with many choices for $n(r)$ leading to the same, exact
free energy.  This complication notwithstanding, the free energy and
$N_\alpha(r)$ obtained at the minimum are meaningful and represent the
exact free energy and solvent site distribution of the combined system
at equilibrium.  (See Ref.~\cite{Petrosyan:2007} for a fuller
discussion.)  Finally, we note that the exact universal functional
conveniently decomposes into a sum of three terms,
\begin{equation} A[n,\{N_\alpha\}]=A_{\textrm{HK}}[n]+
  \Phi_{\textrm{lq}}[\{N_\alpha\}]+\Delta A[n,\{N_\alpha\}], \label{eq:JDFT}
\end{equation}
where $A_{\textrm{HK}}[n]$ is the electronic Hohenberg-Kohn
free-energy functional for the explicit system in isolation,
$\Phi_{\textrm{lq}}$ is the free-energy functional for the liquid when
in isolation, and $\Delta A$ represents the coupling between the
explicit system and the solvent.  Note that, essentially being the
definition of $\Delta A[n,\{N_\alpha\}]$, Eq.~(\ref{eq:JDFT}) is
exact.

Although Eq.~(\ref{eq:JDFT}) is exact in principle,
the forms of these three functionals on the right-hand side are
unknown and need to be approximated in practice.  Generally, such
approximations will break the aforementioned degeneracy and select a
specific $n(r)$ at the minimum of the approximated functional.  In
this {\em Communication}, we use diffusion Monte Carlo to describe the
term $A_{\textrm{HK}}[n]$,
and, for our quantum Monte Carlo implementation of joint density-functional theory, we write the terms related to the environment as 
\begin{equation}
A_{\textrm{env}}[n] \equiv \min_{\{N_\alpha\}}{ \left( \Phi_{\textrm{lq}}[\{N_\alpha\}]+\Delta A[n,\{N_\alpha\}] \right)}.
\end{equation}

The variational derivative of Eq.~(\ref{eq:JDFT}) with respect to the exact electron density then yields the Euler-Lagrange equation $0=\delta A/\delta n=\delta(A_{HK}+A_{\textrm{env}})/\delta n$, which is precisely the equation for the isolated solute system in an external potential $V_{\textrm{env}}\equiv \delta A_{\textrm{env}}[n]/\delta n$.  When an external potential is found for which the electron density $n(r)$ apportioned to the
solute yields back the same potential through this definition, the self-consistent, exact thermodynamic state of the system will have been found.  In principle, this exact solution can be obtained through multiple self-consistent iterations. The main difficulty of a quantum Monte Carlo implementation of $A_{\textrm{HK}}$ in this process is the presence of statistical noise in the resulting electron densities, particularly in regions of low electron density, a problem which also plagues the approach of Ref.~\cite{Amovilli:2008p495}.  There has been recent progress in reducing this noise
\cite{Assaraf:2007p518}, but the results are not yet sufficiently clean for use in self-consistent calculations. We therefore now introduce a method to {\em estimate} the self-consistent solution of our solvation theory {\em without} evaluation of the electron density within quantum Monte Carlo.

\textit{First-order estimator of self-consistency.}
A natural starting point for a joint density-functional theory quantum Monte Carlo calculation is a fully self-consistent joint density-functional theory calculation within the local density (LDA) or generalized gradient (GGA) approximation for $A_{\textrm{HK}}$.  Such calculations provide trial wavefunctions for quantum Monte Carlo calculations, and the resulting densities $n_{\textrm{DFT}}$ give estimates for the environment potential $\tilde{V}_{\textrm{env}}=\delta A_{\textrm{env}}[n_{\textrm{DFT}}]/\delta n$.  One can then perform a single quantum Monte Carlo calculation of the solute within this estimated potential, yielding both an initial quantum Monte Carlo density $n_{\textrm{QMC}}$ and energy $A_{\textrm{QMC}}[n_{\textrm{QMC}}]$.  Note that the latter quantity is defined as the quantum Monte Carlo estimate of the functional $A_{HK}$; i.e., the energy of the solute {\em without} the energy associated with $\tilde{V}_{\textrm{env}}$.  

The two calculations described in the above paragraph could then be combined to give a zeroth-order estimate of the free energy of the solvated system as $A_{\textrm{QMC}}[n_{\textrm{QMC}}]+A_{\textrm{env}}[n_{\textrm{DFT}}]$.  This, however, loses the benefits of the variational principle because the two terms are not self-consistent in that they are evaluated at different electron densities, and so the resulting estimate is not necessarily an upper bound for the final, converged result.   Also, the resulting error is first-order in the errors in the density, rather than second-order, as is usually associated with variational calculations.
We can correct for this by evaluating, with errors only in the second-order, $A_{env}[n_{\textrm{QMC}}]=A_{env}[n_{\textrm{DFT}}]+\int d^3r\, \tilde{V}_{env}\, (n_{\textrm{QMC}}-n_{\textrm{DFT}})$ because, by definition, $\tilde{V}_{\textrm{env}} = \delta A_{\textrm{env}}[n_{\textrm{DFT}}]/\delta n$.  The final result equals
\begin{flalign} 
 \label{eq:secondOrder}
A = &A_{\textrm{QMC}}[n_{\textrm{QMC}}]+A_{\textrm{env}}[n_{\textrm{DFT}}]+ \\
&\int d^3r \tilde{V}_{\textrm{env}}(r) \left( n_{\textrm{QMC}}(r)-n_{\textrm{DFT}}(r) \right) +{\cal O}(\delta n^2), \nonumber   
\end{flalign}
with errors that are second-order in {\em both} the difference between
the exact converged density and the density-functional theory density
$n_{\textrm{DFT}}$ {\em and} the difference between the exact
converged density and the first-iteration density $n_{\textrm{QMC}}$.
Below, we show that this new approach is operationally nearly as good
as full self-consistency, but with the effort of only a single quantum
Monte Carlo calculation.
 
\begin{figure}
\begin{minipage}[c]{85pt}
\includegraphics[width=85pt]{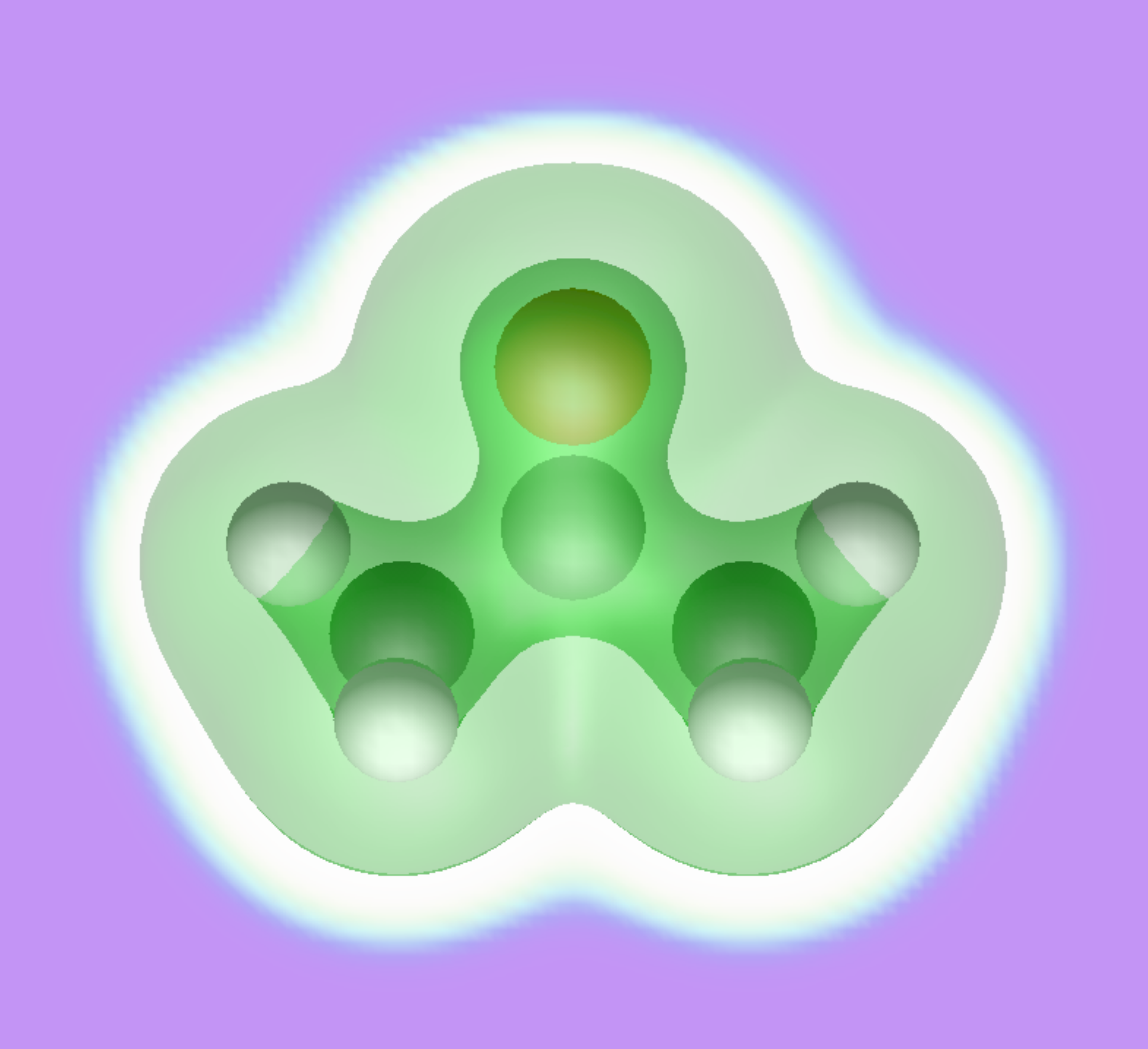}\\(a) 
\end{minipage}
\begin{minipage}[c]{135pt}
\includegraphics[width=135pt]{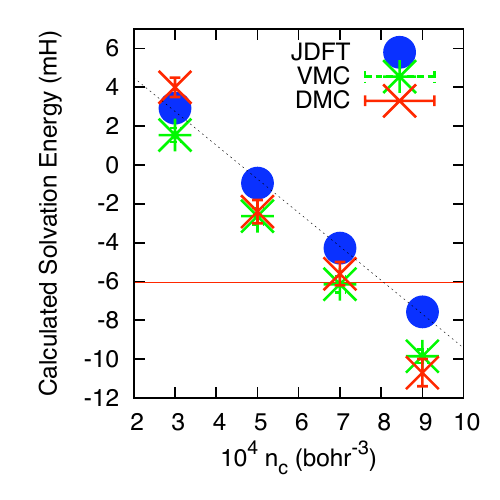}\\(b)
\end{minipage}
 \caption{(Color Online) (a) Joint density-functional theory description of acetone in aqueous solution: electron density contours as green (gray) surfaces, solvent as solid purple (dark gray).  (b) Solvation energies for acetone: diffusion Monte Carlo (DMC) as red (gray) x's, variational Monte Carlo (VMC) as green (gray) strikethrough x's, Joint Density Functional Theory (JDFT) as blue (gray) circles, Joint Density Functional Theory best fit as a dashed black line, and experiment as a horizontal red (gray) line. } \label{acetone}
\end{figure}

\textit{Implementation.}  In this work we demonstrate our solvation technique with a simplified description of the environment, but we stress that it may be employed just as readily with whatever functionals for the liquid become available.  We note that the quality of our results now depend on our approximation for $A_{env}$, which we hope to improve in the future.  Specifically, we here employ the isodensity model of Petrosyan \textit{et al.} \cite{Petrosyan:2005p493}, which is similar to the dielectric model by Fattebert and Gygi \cite{Fattebert:2003p497}.  These models take the dielectric constant to be a local function of the electron density that switches smoothly from the value for vacuum at high solute electron densities to the dielectric constant of the bulk liquid for low electron densities.  This simplified joint density-functional is similar in spirit to the smoothly transitioning nonlinear polarizable continuum model described in Ref. \cite{Amovilli:2006p491}, but here we work in a rigorous framework and also effectively achieve solute-solvent self-consistency.  We have implemented our method in the open source code JDFTx~\cite{jdftx}, interfaced with CASINO~\cite{Needs:2008p339}.  In anticipation of applying our method to large solvated surfaces, we use for our starting point the computationally inexpensive local-density approximation at full solute-solvent self-consistency.  Fig.~\ref{acetone}(a) illustrates the cavity formation for acetone, a small molecule chosen for its well-known experimental solvation energy.  Our reported solvation energies include, in addition to the electrostatic components from our simplified electrostatic functional, the cavitation energy shown in Table~\ref{CavTable}, estimated from classical density functional theory following the procedure in Ref.~\cite{2011arXiv1112.1442S} as included in the JDFTx package~\cite{jdftx} and employing a functional based on Ref. \cite{jefferyAustin}.  

The density-functional theory and Hartree-Fock calculations employ
pseudopotentials from Burkatzki \textit{et al.}
\cite{Burkatzki:2007p451} and expand the wave functions in a
plane-wave basis with a cutoff energy of 30~H (hartree).  The local
density approximation is employed for the density functional theory
calculations, performed using JDFTx~\cite{jdftx}, and a simulation box
of 40 ${\mathrm{bohr}^3}$.  The quantum Monte Carlo calculations are
performed using CASINO~\cite{Needs:2008p339} and employ a trial
wavefunction of a product of a single Slater determinant of density
functional orbitals and a Jastrow correlation factor composed of
electron-nucleus and electron-electron terms with expansion order
eight, and electron-electron-nucleus terms with expansion order three,
as described in Ref.~\cite{jastrow}.  The orbitals and external
potential $V_{\mathrm{env}}$ are represented by B-splines.  The
parameters of the Jastrow factor are optimized by variance
minimization~\cite{VarMin}.  The diffusion Monte Carlo time step is
0.01 H$^{-1}$.  (Going from 0.01 to 0.001 H$^{-1}$, we found the
time-step error to be within the statistical uncertainty we report for
the solvation energy for acetone in Figure~\ref{acetone}, with the solvation parameter $\mathrm{n_c}=7$~x~$10^{-4}~\mathrm{bohr}^{-3}$.)  Molecular geometries are from the Computational Chemistry Comparison and Benchmark Database:  either from experimental data if available, or density-functional optimization using the B3LYP functional and the cc-pVTZ basis set~\cite{NIST}.

\begin{figure}
\includegraphics[width=0.3\textwidth ]{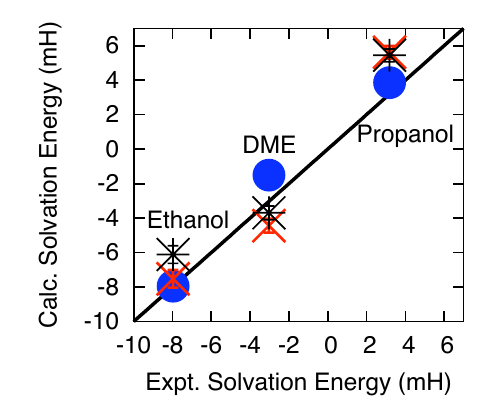} \caption{(Color Online) Theoretical versus experimental solvation energies for ethanol, dimethyl ether (DME), propane (left to right): Joint Density Functional Theory as blue (gray) circles, zeroth-order diffusion Monte Carlo as black strikethrough x's, first-order corrected diffusion Monte Carlo as red (gray) x's.  Experimental values from \cite{Barone:1997p507}.  } \label{multiplot}
\end{figure}

\begin{table}[t]
\begin{tabular} { | l | c | c | c | c | c | c }
\hline
 $ 10^{4}$~$\mathrm{n_c}$~($\mathrm{bohr}^{-3}$)  & 3.0 & 5.0 & 7.0 & 8.1 & 9.0 \\
  \hline  
 dimethyl ether & 9.0 & 7.8  & 7.0 &6.7 & 6.5 \\
  ethanol & 8.9 &   7.7 &  7.0 &  6.7 & 6.4 \\
 propane & 10.3 & 8.9 & 8.0 & 7.7 &  7.5 \\
acetone & 10.6 & 9.2 & 8.3 & 8.0 & 7.7 \\
\hline 
\end{tabular}
\caption{Formation energies (mH) for cavities at various values of the electron isodensity-contour parameter $n_c$ from the model of Ref.~\cite{Petrosyan:2005p493}, based on the cavitation energy method of Ref.~\cite{2011arXiv1112.1442S}. }
\label{CavTable}
\end{table}


For the continuum description of water, we use the local dielectric
function 
$\epsilon(n(r))=1+(\epsilon_\textrm{b}-1)\,\textrm{erfc}\left(
\textrm{ln}(\frac{n(r)}{n_c})/(\sqrt{2}\sigma) \right)/2$,
where $\epsilon_\textrm{b}$ is the bulk dielectric constant of the
fluid, $n_c$ specifies the value of the solute
electron density at which the dielectric cavity forms, and $\sigma$
controls the transition width~\cite{Petrosyan:2005p493}.  This description of water leads to the expression 
$A_{env}[n_{\textrm{DFT}}]=\frac{1}{2}\int dr^3 (n_{\textrm{DFT}}-N)(-4\pi(\nabla \cdot \epsilon(n_{\textrm{DFT}}) \nabla)^{-1}+ 4\pi(\nabla^2)^{-1})(n_{\textrm{DFT}}-N)$, where $N$ represents the nuclei of the solute.

\textit{Results.} Figure~\ref{acetone}(b) shows the resulting solvation energies for acetone as a function of $n_c$.  The results are not sensitive to $\sigma$, so we leave this value fixed at 0.6 and optimize the values of $n_c$ for use in diffusion Monte Carlo.  All of the variational Monte
Carlo results and nearly all the diffusion Monte Carlo results lie
below the density-functional theory data.  The variational Monte Carlo
results with no Jastrow factor (and thus no correlation, but
exact exchange energy) and the diffusion Monte Carlo results
lie very near to each other, indicating that the primary corrections
in solvation energy to the density-functional results come from the
exact treatment of the exchange and that corrections to
correlation beyond the local-density approximation largely cancel for the solvation energies, at least for acetone.

%

A least squares fit of the data shown in Fig.~\ref{acetone}b yields $n_c=7.0\times 10^{-4}$~$\mathrm{bohr}^{-3}$ as optimal for diffusion Monte Carlo and $n_c=8.1\times10^{-4}$~$\mathrm{bohr}^{-3}$ as optimal for use with density-functional theory.  Figure~\ref{multiplot} shows the resulting solvation energies of three molecules for both the zeroth-order expression and the first-order corrected version from Eq.~\eqref{eq:secondOrder}.  The agreement between the quantum Monte Carlo results and experiment is encouraging given the particularly simple model employed here for the fluid.  The figure also demonstrates the importance of using Eq.~(\ref{eq:secondOrder}) to include the effects of self-consistency, particularly for ethanol.

\begin{figure}
\includegraphics[width=0.35\textwidth ]{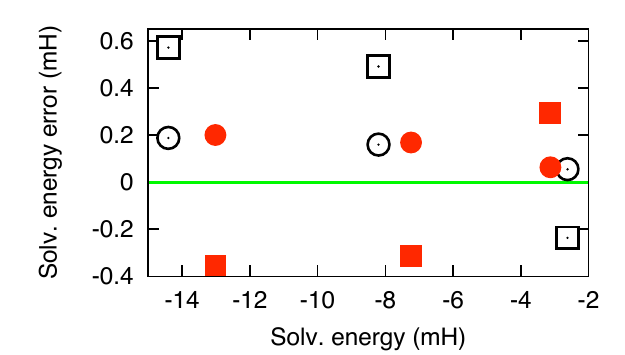} \caption{(Color Online) Zeroth order approximation errors (squares) and first order approximation errors from Eq.~(\ref{eq:secondOrder}) (circles) relative to the exact self-consistent fluid-electronic structure minimization for Hartree-Fock calculations with LDA potentials in black (unfilled) and LDA calculations with Hartree-Fock potentials in red (gray) (filled), for the electrostatic solvation energies of ethanol, dimethyl ether, and propane (from left to right).  } \label{LDAHF}
\end{figure}

To estimate the remaining error between the corrected formula
(Eq.~\eqref{eq:secondOrder}) and full self-consistency, we employ two
different electronic structure methods for which achievement of full
self-consistency is feasible and whose difference in densities we
expect to be similar to the density
difference between density-functional theory and quantum Monte Carlo.
We begin with an environment potential $V_{env}$ from a solvated density-functional theory
(Hartree-Fock) calculation, and include it in a Hartree Fock (density-functional theory) calculation, attempting to predict the final
self-consistent energy using our proposed methods.  Fig.~\ref{LDAHF} compares the zeroth and
first-order corrected approximations with the fully self-consistent
results when working this procedure in both directions.

The data exhibit a number of behaviors which we expect to be general
trends.  First, the first-order corrected data lie above the fully
self-consistent result, regardless of the starting point.  Second, the
remaining (second-order and higher) errors are all quite small (0.2~mH
or less) and on the order of about one-third the size of the
first-order correction.  Third, the remaining errors in the corrected
results for a given molecule when going in either direction between
density-functional theory and Hartree Fock are nearly identical.  The
difference between these remaining errors gives an estimate of the correction at odd orders (third and higher), indicating that the errors
remaining are dominated by the
second-order term.  In aqueous solution, electrostatic screening, a
negative-definite quantity, dominates this second-order term, thus
explaining why the corrected results generally lie above the self-consistent
solution.

These observations strongly suggest that the self-consistency
error in our corrected diffusion Monte Carlo results is less than one
mH for ethanol, and even less for the other molecules.  Our
procedure thus likely gives an upper bound well within chemical
accuracy of the results of full self-consistency, {\em with the need for only a single quantum Monte Carlo calculation}.

\textit{Conclusion.} The framework of joint density-functional theory
provides a rigorous and efficient method for solvation in quantum
Monte Carlo, without the need for phase-space sampling of the fluid.
Moreover, a special procedure allows self-consistency of the joint
calculation to be obtained (along with an estimate of the remaining
errors) to within {\em chemical accuracy}, all at the cost of only a {\em single} quantum Monte Carlo total-energy calculation carried out in
a fixed external potential.  This solvation method applies
just as readily to molecular and surface calculations.  The procedure
is general and may be used not only with quantum Monte Carlo but other
correlated total-energy methods as well.

K.A.S. and K.L.W were supported by National Science Foundation Graduate Research Fellowships, K.A.S. and R.G.H. by Contract No.\ CAREER DMR-1056587, and K.A.S. and T.A.A. by DE-FG02-07ER46432.  R. S. and K.L.W. were supported under Award Number DE-SC0001086. Computational support was provided by the Computation Center for Nanotechnology Innovation at Rensselaer Polytechnic Institute.


%
\end{document}